\documentclass[reprint,aps,pra,two-column,superscriptaddress]{revtex4-1}
\usepackage{bm,amsmath,amssymb,amsfonts}
\usepackage{color}
\usepackage{dcolumn}
\usepackage{graphicx}
\usepackage{float}
\usepackage{hyperref}
\hypersetup{colorlinks=true,citecolor=blue,urlcolor=blue,linkcolor=blue}
\usepackage{ulem}
\usepackage{times}
\usepackage{subfigure}
\usepackage{appendix}

\begin{document}
\title{Quantum phase transitions of the anisotropic Dicke-Ising model in driven Rydberg arrays}
\author{Bao-Yun Dong}
\affiliation{Department of Physics, and Chongqing Key Laboratory for Strongly Coupled Physics, Chongqing University, Chongqing, 401331, China}
\author{Ying Liang}
\affiliation{Department of Physics, and Chongqing Key Laboratory for Strongly Coupled Physics, Chongqing University, Chongqing, 401331, China}
\author{Stefano Chesi}
\thanks{corresponding author: stefano.chesi@csrc.ac.cn}
\affiliation{Beijing Computational Science Research Center, Beijing 100193, China}
\affiliation{Department of Physics, Beijing Normal University, Beijing 100875, China}
\author{Xue-Feng Zhang}
\thanks{corresponding author:  zhangxf@cqu.edu.cn}
\affiliation{Department of Physics, and Chongqing Key Laboratory for Strongly Coupled Physics, Chongqing University, Chongqing, 401331, China}
\affiliation{Center of Quantum Materials and Devices, Chongqing University, Chongqing 401331, China}

\begin{abstract}
We study the properties of a generalized Dicke-Ising model realized with an array of Rydberg atoms, driven by microwave electric fields and coupled to an optical cavity. As this platform allows for a precisely tunable anisotropy parameter, the model exhibits a rich landscape of phase transitions and critical phenomena, induced by the interplay of rotating-wave, counter-rotating-wave, and Ising interactions. We develop an improved quantum Monte Carlo algorithm based on the stochastic series expansion that explicitly tracks the Fock state of the quantum cavity. In the superradiant (SR) phase, this allows us to determine, through data collapse, the scaling laws of the photon number. We also demonstrate the vanishing of parity symmetry in finite-size simulations and show that the Rydberg blockade leads to a significant suppression of cavity occupation. Notably, stronger quantum fluctuations induced by the counter-rotating wave terms slightly favor the superradiant solid (SRS) phase over the Solid-1/2 state. Finally, we confirm that the SR phase transition and the transition from the Solid-1/2 to the SRS are second-order. In contrast, the transitions from the Solid-1/2 or SRS to the SR phase are both first-order for any value of the normalized anisotropy parameter. 
\end{abstract}
\maketitle

\section{INTRODUCTION}

The quantum Rabi model (QRM) \cite{WOS:000542185800009, PhysRev.49.324, PhysRev.51.652, Zhong_2013, Xie_2017} and the Dicke model (DM) \cite{PhyDicke, Nahmad-Achar_2013, WOS:001157079100006, PhysRevLett.113.023603, PhysRevLett.127.253601} are fundamental frameworks for understanding the interaction between atoms and quantized light. In particular, as discussed by several theoretical and numerical analysis~\cite{PhyDicke, PhysRevE.67.066203, PhysRevX.4.021046, PhysRevLett.117.123602, PhysRevLett.115.180404, PhyMott, Baumann, Liu, PhysRevA.108.063716}, they have been instrumental in predicting various types of superradiant phase transitions (SRPT), both in few-body and many-body systems. The SRPT is a transition from the normal phase (NP) to a superradiant phase (SRP), characterized by a macroscopic number of photons. The transition is caused by quantum fluctuations, which lead to a spontaneous breaking of the $Z_{2}$ symmetry. 

In the scenario of weak coupling between matter and light, the rotating wave approximation (RWA) is valid~\cite{1443594}. However, as the coupling strength approaches the atomic transition frequency, i.e., the system enters the strong coupling regime,  effects of the counter-rotating-wave (CRW) terms become comparable to those of the rotating-wave (RW) terms, rendering the RWA inapplicable~\cite{Yuyi}. The presence of CRW terms also leads to a reduction of the continuous $U(1)$ symmetry of the Jaynes-Cummings model \cite{Meystre2021} to a discrete $Z_{2}$ symmetry. While in the conventional Dicke model RW and CRW terms have equal strength, the study of a generalized DM, also named anisotropic Dicke model (ADM), has recently sparked considerable interest~\cite{Das063716, KLOC201785, Aedo042317, Shapiro023703}. The defining feature of this ADM is a difference in the coupling strengths of RW and CRW interactions. The ratio
between the CRW coupling strength and the total coupling strength (CRW + RW) can be defined as a normalized anisotropy parameter (NAP).
The ADM exhibits not only the phase transition from NP to SRP, but also a transition from an ergodic to a nonergodic phase at finite temperature~\cite{Buijsman08060, Hu_2021, Das043706}.

The observation of a SRPT has been reported in various experiments based on ultra-cold atoms~\cite{Baumann, PhyMott, Keeling}. Furthermore, the ADM could be realized with superconducting quantum circuit architectures~\cite{Nataf}. 
Rydberg arrays interacting with an optical cavity represent another promising way to investigate the SRPT, in the presence of the additional physical mechanism associated with the Rydberg-Rydberg interactions~\cite{Guerlin53832, Mivehvar02012021,exp1,exp2,exp3,exp4}.
The corresponding model, describing the interplay between strong light-matter and matter-matter interactions, is known as the Dicke-Ising Hamiltonian ~\cite{Rohn31, Cortese96, Jan004, Nevado3624, Puel106901}.
In the antiferromagnetic case, the competition between these two interactions, which can break the translational or $Z_{2}$ symmetry, gives rise to a rich phase diagram.
While the atom-light interaction favors the formation of a SRP, the Ising interaction tends to form Rydberg crystals, e.g., inducing a transition between NP and a Solid-1/2 (antiferromagnetic) phase.
Most importantly, several studies of this model have consistently reported a superradiant solid  (SRS) phase, in which solid order and superradiance coexist~\cite{Zhang24522250, zhang01, AnGaoqi, lAnji202411}. It was also shown that, after performing the RWA on the Dicke-Ising Hamiltonian, a geometrically frustrated triangular Rydberg array interacting with quantized light gives rise to a novel superradiant clock phase~\cite{liang2025}. Furthermore, the non-equilibrium phenomena in the cavity-coupled Rydberg arrays are also exotic \cite{Prethermalization,Marcello}

In this paper, we investigate in detail the properties of the anisotropic Dicke-Ising model (ADIM). Motivation for our study arises from a recent proposal based on the cavity-Rydberg array platform, showing that the anisotropy parameter can be controlled precisely when the cavity-Rydberg array is driven by microwave electrical fields~\cite{dongade388}. 
In a realization of such ADIM, the phase transitions and critical behavior are expected to be particularly rich, being determined by the interplay of RW, CRW, and Ising terms. Meanwhile, the influence of $U(1)$ symmetry breaking to $Z_2$ symmetry on the SRS phase and phase transition can also be figured out.

To study this system, we first apply the mean-field approximation, in which the light-matter Hamiltonian is reduced to a transverse-field Ising model independent of the NAP. 
Afterwards, we explore the ADIM quantitatively by developing a specific cluster stochastic series expansion (SSE) algorithm~\cite{Sandvik1, Sylju_sen_2002, DXLiuL140404}, which is a type of quantum Monte Carlo technique. 
Specifically, since the traditional directed loop update fails for a general spin-boson system in the presence of CRW terms, the wormhole algorithm was previously applied to the DIM after integrating out the bosonic field~\cite{lAnji202411}. Here, instead, we have redesigned the directed loop update process to explicitly include the Hilbert space of the quantum cavity, and we have optimized the transfer matrix to arbitrarily large dimensions.
Therefore, we can extract the photon scaling laws of the ADM by employing data collapse in the SSE simulation. We also observe that the parity vanishes in the SR phase for any large but finite system size. 
In the SR phase of the ADIM, we find that the Rydberg blockade leads to a substantial reduction of the cavity occupation.
Furthermore, we provide evidence of stronger quantum fluctuations induced by the CRW terms, compared to the RW interaction, by showing that a larger NAP moves the phase boundary between the SRS and Solid-1/2 states. Besides confirming that the transition from the NP to the SR phase, as well as that between Solid-1/2 and SRS, are second-order, we find that the transitions from the Solid-1/2 or SRS phases to the SR phase are first-order for arbitrary values of the NAP.

The paper is organized as follows:
In Sec. \ref{MEAN-FIELD}, we discuss the model Hamiltonian and its mean-field approximation.
In Sec.~\ref{SSE} , we introduce our improved SSE approach.
Section~\ref{numerical1} examines the behavior of the cavity photon and parity in the ADM around the SRPT point.
Section~\ref{numerical1} is dedicated to analyzing the rich phase diagram of the ADIM.
In Sec.~\ref{outlook} we present our conclusions and outlook.

\section{Model and Mean-field analysis}
\label{MEAN-FIELD}

\begin{figure}
	\centering
	\includegraphics[width=1.0\linewidth]{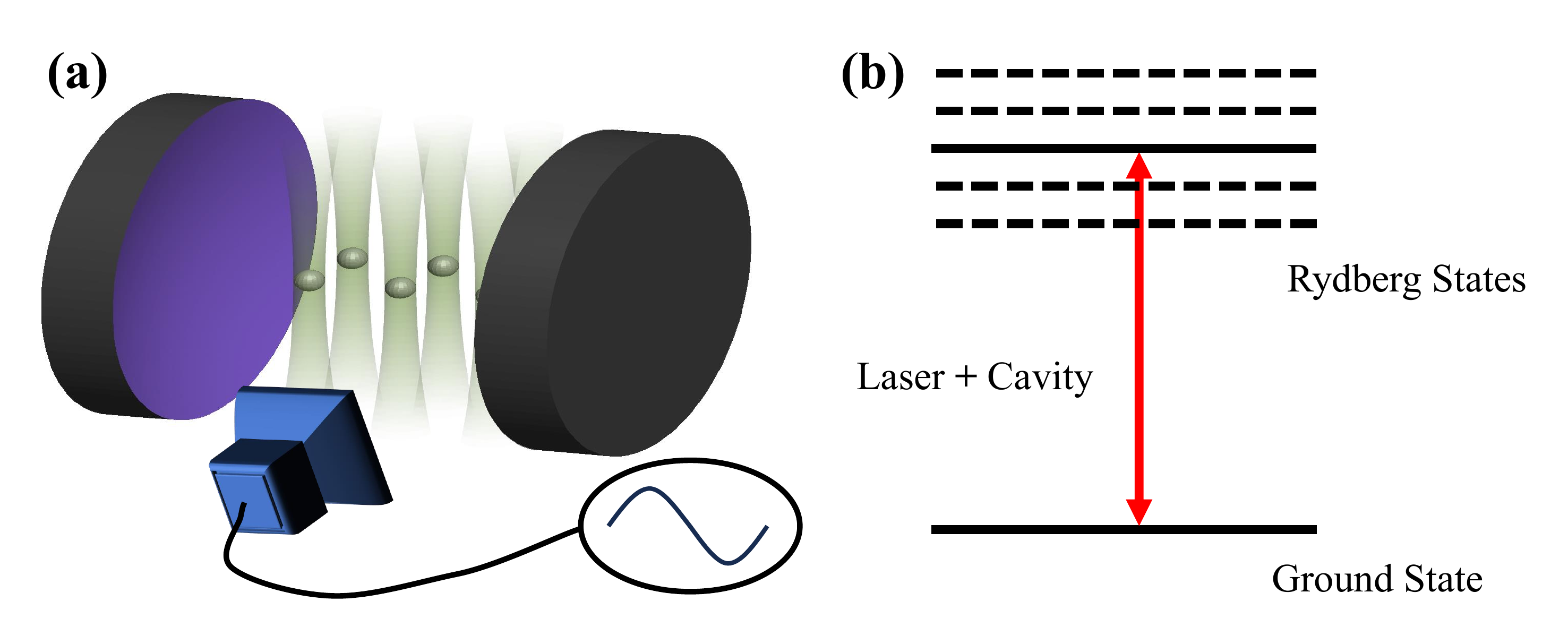}
	\caption{(a) Illustration of the proposed setup, with a two-dimensional Rydberg atom array coupled to an ultrahigh finesse optical cavity and driven by a microwave electric field. 
    (b) Schematic energy-level scheme of a driven Rydberg atom. } 
    \label{fig1}
\end{figure}

We consider $\rm ^{87}Rb$ atoms trapped by optical tweezers and interacting with both an ultrahigh finesse optical cavity and a classical drive. The optical tweezers are arranged to form a two-dimensional square lattice, and the system is driven by a microwave electrical field with the overall experimental setup illustrated in Fig.~\ref{fig1}(a).  As a consequence of the drive, the Rydberg excitation splits into a series of Floquet states, schematically shown in  Fig.~\ref{fig1}(b), with energy separation determined by the microwave modulation frequency. 
In such a setup, it has been shown  that the following effective Hamiltonian can be realized~\cite{dongade388}, where a bosonic field is coupled to an ensemble of hard-core bosons with repulsive interaction:
\begin{align}
    H & = V\sum_{\langle j,k \rangle} n_{j} n_{k}
    - \mu a^{\dagger}a  - (\mu + \Delta)\sum_{j=1}^{N} n_{j} \nonumber\\
    & + (1-\alpha) \frac{g}{\sqrt{N}} \sum_{j=1}^{N} ( \sigma_{j}^{+} a +  a^{\dagger} \sigma_{j}^{-}) \nonumber\\
    &+ \alpha \frac{g}{\sqrt{N}} \sum_{j=1}^{N} ( \sigma_{j}^{+} a^{\dagger} +a \sigma_{j}^{-}) . \label{Ohamil}
\end{align}
Here, the first term is a nearest-neighbor Ising-like repulsive interaction with coupling strength $V$, where $n_{j} =\frac12 (\sigma_{j}^z + 1)$ is the Rydberg occupation operator. We use the notation $\sigma_{j}^{\gamma}$, where $\gamma \in \left\{ x,y,z \right\}$, to denote the Pauli matrices associated with different sites $j = 1,2\ldots N$. Then, the operators $\sigma_{j}^{\pm} = \sigma_{j}^{x} \pm i\sigma_{j}^{y}$ can describe the transitions between the atomic ground states and the Rydberg states. Throughout this article, we will consider a square lattice with $N=L^2$ sites and periodic boundary conditions. The second and third terms of Eq.~(\ref{Ohamil}) represent the chemical potentials of the cavity photon and of the Rydberg state, given by  $\mu$ and $\mu + \Delta$, respectively.  The light field is described by a bosonic annihilation (creation) operator $a$~$ (a^{\dagger})$, while $\Delta$ is the detuning between the cavity and the Rydberg atoms, which is easily tunable in the cavity-Rydberg platform. The last two terms of Eq.~(\ref{Ohamil}) arise from the light-matter interaction. The collective coupling $g$ is proportional to both the strength of the classical drive and the bare coupling to the cavity mode~\cite{dongade388}, while $\alpha$ controlling the degree of anisotropy. This NAP $\alpha$ determines the relative strength between the RW terms (annihilating a cavity photon and exciting an atom, and vice versa) and CRW terms (simultaneously exciting or annihilating the cavity and one of the atoms).
Experimentally, the periodic microwave drive can realize a continuously tunable NAP in the range $\alpha \in [0,1]$.

We now perform a mean-field approximation by treating the bosonic operators as classical light fields.
Explicitly, we consider a state of the form $| \Psi \rangle \simeq |\Phi \rangle \otimes | \lambda \rangle$, where $|\Phi \rangle$ is a generic atomic state and $| \lambda \rangle = \exp{(-\frac{1}{2}|\lambda|^2)} \mathrm{e}^{\lambda a^{\dagger}} | 0 \rangle$ is a coherent state of the bosonic photon field, with $|\lambda|^{2} \propto N$. This approximation, which neglects correlations between the light field and the atomic state, is expected to be accurate in the SRP when the cavity is populated by a large number of photons. On the other hand, many-body correlations within the atomic system are still taken into account. In fact, as we will see, $| \Phi \rangle $  is the ground state of an interacting Hamiltonian.
Following this partial mean-field approximation, we compute $H_\mathrm{M} = \langle \lambda | H | \lambda \rangle$, giving:
\begin{align}
    H_\mathrm{M} &= V \sum_{\langle j,k\rangle} n_{j} n_{k} - \mu|\lambda|^{2} - (\mu + \Delta) \sum_{j=1}^{N} n_{j}\nonumber\\
    &+ \frac{g}{\sqrt{N}}\sum_{j=1}^{N} \left[ \lambda_{R}\sigma_{j}^{x} + (1 - 2\alpha) \lambda_{I}\sigma_{j}^{y} \right],
    \label{smpli}
\end{align}
where $\lambda_{R}={\rm Re}[\lambda]$ and $\lambda_{I}={\rm Im}[\lambda]$. 

To bring Eq.~(\ref{smpli}) to a more transparent form, we apply a rotation in the $x-y$ plane, $U = \exp{ \left( i \frac{\theta}{2}\sum_{j=1}^{N}\sigma_j^z \right)}$ with $\theta = \arcsin{\left( \frac{\lambda_{R}}{\sqrt{\lambda_{R}^{2} + (1-2\alpha)^{2} \lambda_{I}^{2}}} \right)}$. The transformed Hamiltonian, $H_\mathrm{MR} = U^{\dagger} H_\mathrm{M}U$, reads as follows:
\begin{align}
    H_\mathrm{MR} &= V\sum_{\langle j,k \rangle} n_{j} n_{k} 
    - \left( \mu + \Delta \right) \sum_{j=1}^{N} n_{j}+ h \sum_{j=1}^{N} \sigma_{j}^{x}-\mu |\lambda|^2.
    \label{MR}
\end{align}
As seen, $U$ does not change the diagonal part of the Hamiltonian, but the atom-cavity interaction induces a transverse field along $x$, with $h$ given by:
\begin{equation}
h = \frac{g}{\sqrt{N}}\sqrt{\lambda_{R}^{2}+(1-2\alpha)^{2}\lambda_{I}^{2}}.
\end{equation}
Since the cavity field enters the atomic part of the Hamiltonian only through $h$, it is useful to parameterize the coherent state with $(h,\lambda_I)$, instead of $(\lambda_R,\lambda_I)$. The mean number of cavity photons $ n_{\rm ph}=|\lambda|^2 $ becomes:
\begin{equation}
n_{\rm ph} = \frac{N h^2}{g^2} + 4 \alpha (1-\alpha)\lambda_I^2,
\end{equation}
which allows us to immediately conclude, using $\alpha(1-\alpha)\geq 0$ and $\mu<0$, that the total energy is minimized at any given $h$ by choosing $\lambda_I = 0$. The ground state of $H_M$ may then be found by considering, equivalently, the following Hamiltonian:
\begin{align}
    H_\mathrm{IM} = & V\sum_{\langle j,k \rangle} n_{j} n_{k}
    - \left( \mu + \Delta \right) \sum_{j=1}^{N} n_{j} \nonumber \\
   & + g\sqrt{\frac{n_{\rm ph}}{N}}\sum_{j=1}^{N} \sigma_{j}^{x} - \mu n_{\rm ph},
    \label{IM}
\end{align}
which has the notable feature of depending on the number of photons $n_{\rm ph}$, but not on the NAP $\alpha$.
 
\begin{figure}
    \centering
    \includegraphics[width=1.0\linewidth]{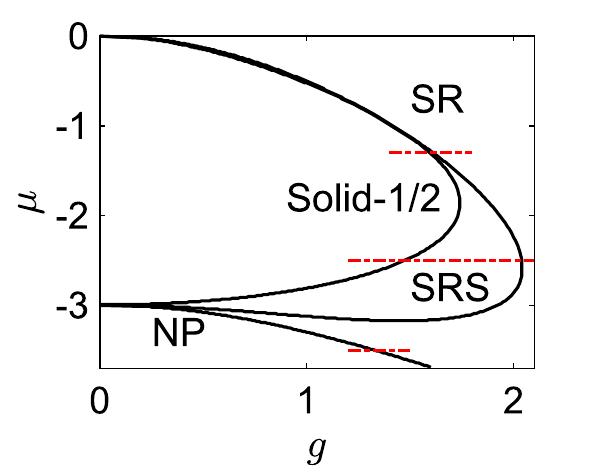}
    \caption{Quantum phase diagram of the AIDM, obtained using the variational state of Eq.~(\ref{VariationalMF}). All phase boundaries are independent of the NAP $\alpha$. The bottom, middle, and top (red dashed) line cuts are analyzed in Fig.~\ref{phsupress}, Fig.~\ref{solid}(c,d), and Fig.~\ref{solid}(a,b), respectively. The other parameters are $\Delta = 3$ and $V = 1$.}
    \label{ADMdata}
\end{figure}

The arguments above show that, within mean-field theory, the ground state of this cavity-Rydberg array is independent of $\alpha$. Therefore, if the light field can be treated classically, the system may be characterized through previous studies of the Dicke-Ising Hamiltonian without anisotropy. In particular, our conclusions are consistent with the analysis of anisotropic Dicke and Rabi models, where the $V$ is absent~\cite{Nataf,Liu}. In the presence of Rydberg-Rydberg interactions, the atomic many-body problem translates to the study of an Ising model with both longitudinal and transverse fields, see Eq.~(\ref{IM}). By also neglecting atomic correlations, we can  obtain an approximate description based on the following trial state~\cite{zhang01}:
\begin{align}
    |\Psi (\lambda, \theta_A, \theta_B) \rangle 
    = \bigotimes_{i \in A } \left ( \cos\frac{\theta_{A}}{2} |\! \uparrow_{i} \rangle + \sin\frac{\theta_{A}}{2}|\! \downarrow_{i} \rangle\right) \nonumber\\
    \bigotimes_{j \in B }\left ( \cos\frac{\theta_{B}}{2} |\! \uparrow_{j} \rangle + \sin\frac{\theta_{B}}{2}|\! \downarrow_{j} \rangle\right) \otimes |\lambda\rangle, 
    \label{VariationalMF}
\end{align}
where the square lattice has been partitioned into two sublattices, $A$ and $B$, to capture the effects of the antiferromagnetic interaction. An example of the phase diagram, obtained by minimizing the energy over the variational parameters $\lambda$ and $\theta_{A,B}$, is shown in Fig.~\ref{ADMdata}. The superradiant phases (SR and SRS) are characterized by $\lambda \neq 0$, while the phases breaking translational symmetry (Solid-1/2 and SRS) have $\theta_A \neq \theta_B$.

Beyond the mean-field theory, a second-order all-to-all spin-exchange interaction can emerge among the atoms mediated by the cavity. Furthermore, the CRW terms, controlled by the NAP $\alpha$, enhance quantum fluctuations. Given the expected significant impact of these effects -- particularly in finite-size systems where mean-field theory can fail -- we have developed a quantum Monte Carlo algorithm based on the SSE approach, which allows us to investigate the exact ground state of this model accurately.

\section{Quantum Monte Carlo method}
\label{SSE}

Before presenting the new formula for updating off-diagonal operators, we first review the implementation of the SSE algorithm in photon-atom coupled systems.
The core idea of the algorithm is to expand the partition function into a series and then use Markov Chain Monte Carlo sampling to numerically study the quantum system.
In the absence of a sign problem, the computational complexity of the directed-loop SSE algorithm is proportional to the system size and the inverse temperature.
In high-dimensional systems, with a nonlocal coupling of the bosonic field to the atoms, the SSE algorithm exhibits higher computational efficiency than alternative methods, e.g., based on matrix product states (which require high bond dimensions).
However, the nonlinear CRW terms violate local particle number conservation, making
impossible to implement the conventional directed-loop algorithm.

As a consequence, previous implementations of the SSE algorithm were based on wormhole updates after having integrated out the bosonic field, via path integrals in the coherent state representation. This approach avoids the need for photon sampling and improves algorithmic efficiency.
Instead, similar to the SSE method for the spin XYZ model \cite{kagome_nc}, here we have developed a directed-loop algorithm that also updates the CRW operators, which don't conserve the total density.
As we will discuss, direct sampling of the bosonic field can provide additional and physically relevant information about the properties of the various phases arising from the model.

As in the conventional SSE, we start by expanding the partition function into a series:
\begin{align}
    Z = \sum_{| \Psi \rangle} \sum_{k=0}^{\infty} 
    \frac{(-\beta) ^{k}}{k !} \langle \Psi | \prod_{p=1}^{k} H_{a_{p}, b_{p}} |\Psi\rangle,
\end{align}
where $\beta = \frac{1}{k_{B}T}$ is the inverse temperature. We evaluate the partition function using the Fock basis, instead of relying on a coherent state representation of the cavity mode state. Therefore, the set of basis states (equivalent to configurations) is chosen as $|\Psi\rangle =\left( \bigotimes_{j=1}^{N}  | \sigma_{j} \rangle \right) \otimes |m\rangle$, where $|\sigma_j\rangle = |\!\uparrow_j\rangle,|\!\downarrow_j\rangle$ are the eigenstates of $\sigma_j^z$. We decompose the Hamiltonian Eq.~\eqref{Ohamil} into local operators $H_{a_{p}, b_{p}}$, which act on the configuration without creating superpositions. 
Here, $p$ denotes the position in imaginary time, $a$ represents the operator type (diagonal and off-diagonal operators, labeled as $d_\gamma, rw_\gamma^\pm, cw_\gamma^\pm$), and $b \in \{ 1, 2, ..., N  \}$ indicates the position in real space where the operator acts. Notice that different types of decomposition can be selected depending on the geometry of the clusters.
Here, we choose three nearest neighbor Rydberg sites in a line plus the photon site as a whole cluster,  so the diagonal operators at the site $j$ are explicitly written as:
\begin{align}
    H_{d_{\gamma}, j} &=C_{1}V \left( n_{j} n_{k} + n_{j} n_{l} \right) 
    - C_{2}\mu a^{\dagger}a \nonumber\\
    &- C_{3}(\mu + \Delta)(n_{j} + n_{k} + n_{l})
    \label{DO}
\end{align}
where $k$ and $l$ label the two nearest-neighbor sites in the direction  $\gamma$. For a square lattice, $\gamma=1$ and $\gamma=2$ correspond to the horizontal and vertical directions, respectively.  $C_{1}, C_{2}, C_{3}$ are correction factors to avoid overcounting, and their values are not unique. Here, we simply choose $C_{1} = 1/2$, $C_{2} = 1/(2N)$, and $C_{3} = 1/6$ in a square lattice with $N$ atoms. On the other hand, the corresponding off-diagonal operators are:
\begin{align}
    H_{rw_{\gamma}^+, j} &= (1-\alpha) C_{4} \frac{g}{\sqrt{N}}\sigma_{j}^{+}a, \nonumber\\
    H_{rw_{\gamma}^-, j} &= (1-\alpha) C_{4} \frac{g}{\sqrt{N}} a^{\dagger} \sigma_{j}^{-},\nonumber\\
    H_{cw_\gamma^+, j} &= \alpha C_{4} \frac{g}{\sqrt{N}} a^{\dagger} \sigma_{j}^+, \nonumber\\
    H_{cw_{\gamma}^-, j} &= \alpha C_{4} \frac{g}{\sqrt{N}} \sigma_{j}^{-}a.
    \label{off_diag_operators}
\end{align}
Note that these operators do not actually depend on the value of $\gamma=1,2$, which is compensated by the overcounting factor $C_{4} = 1/2$. 
From Eq.~(\ref{off_diag_operators}), we can also see that the operators $H_{cw_\gamma^\pm, j}$ do not conserve the number of excited particles, which causes the conventional directed-loop update to fail in this system. Therefore, while diagonal updates and measurements of physical quantities follow the established SSE algorithm of Ref.~\cite{Sandvik1}, a new type of directed-loop update must be developed for these special operators.

\begin{figure}
    \centering
    \includegraphics[width=1.0\linewidth]{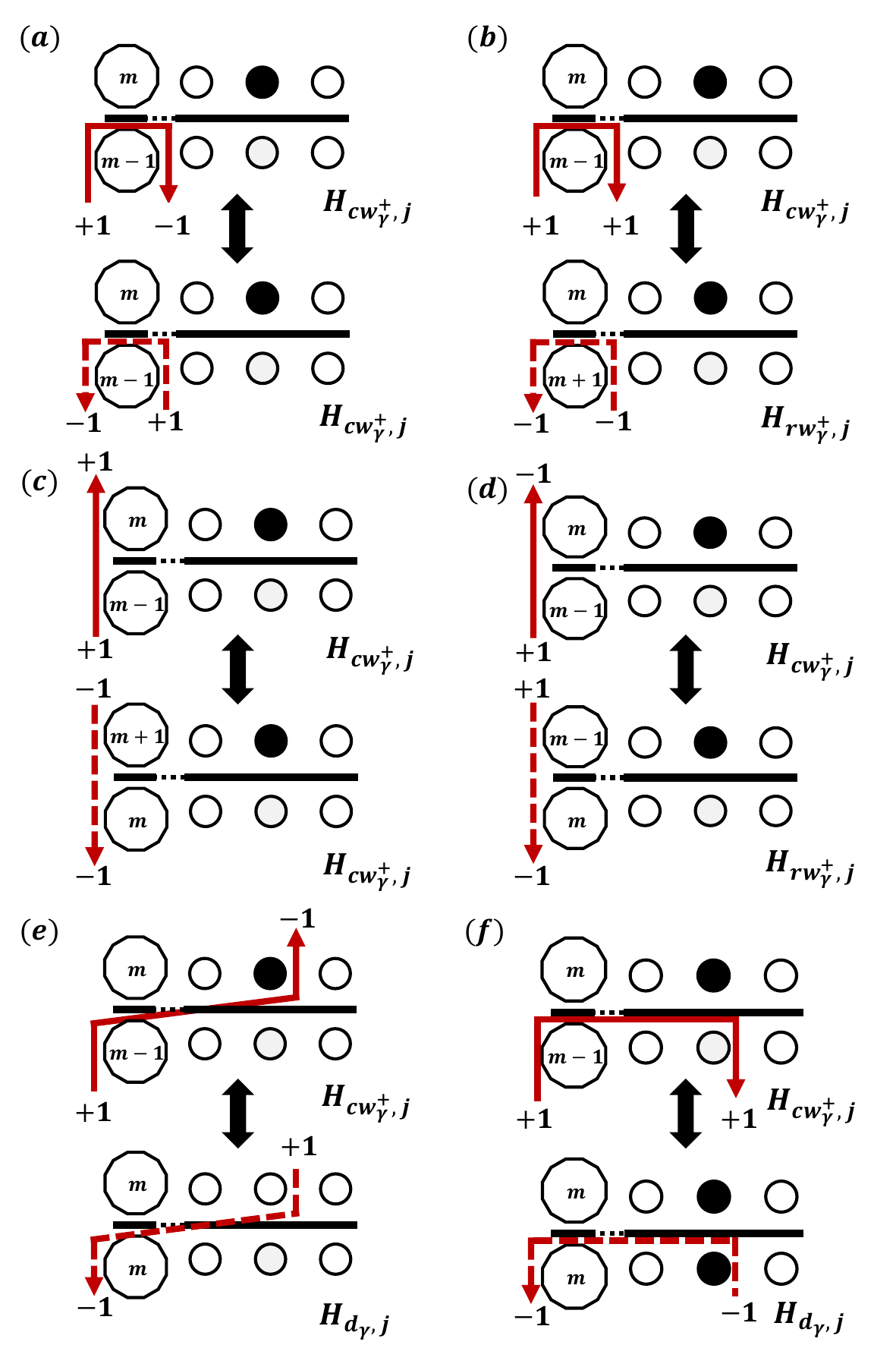}
    \caption{
    Possible assignments of the worm head (red solid-line arrow) entering an $H_{cw_{\gamma}^+, j}$ type vertex from the lower left leg (cavity mode state).
    The numbers in the large polygons indicate the Fock state of the cavity, while black (white) circles represent atoms in the Rydberg (ground) state, the numbers ($\pm 1$) on the arrow head and tail correspond to (add, minus) one of the cavity mode state or flip (up, down) of the atom.
    The worm head is \textit{bounced back} with (a) vertex unchanged or (b) cavity mode state increasing two photons and updating to $H_{rw_{\gamma}^+, j}$; 
    \textit{continue straight} with cavity mode state changed and (c) operator unchanged or (d) updating to $H_{cw_{\gamma}^+, j}$; (e) \textit{cross through} and (f) \textit{turn back} with updating to $H_{d_{\gamma}, j}$. 
    The corresponding inverse process is represented by the red dashed arrows.} 
    \label{vertices}
\end{figure}

As an example, we show in Fig.~\ref{vertices} the update rules for a vertex associated with $H_{cw_{\gamma}^+,j}$, assuming that the worm head enters the vertex from the lower left leg and adding one cavity mode state. Different from the conventional SSE method, here the state at the leg where the worm head exits can be either raised or lowered by one. 
Fig.~\ref{vertices}(b)(d) and Fig.~\ref{vertices}(e)(f) present the case of $H_{rw_{\gamma}^+, j}$ update to $H_{rw_{\gamma}^+, j}$ and $H_{d_{\gamma}, j}$, respectively.
Fig.~\ref{vertices}(c) updates the cavity mode state, although the operator remains unchanged.
In other cases, for example, when the worm head reaches $H_{rw_{\gamma}^-, j}$ or enters from the atomic side for all vertices, the update procedure follows a similar scheme.


\begin{figure}
    \centering
    \includegraphics[width=1.0\linewidth]{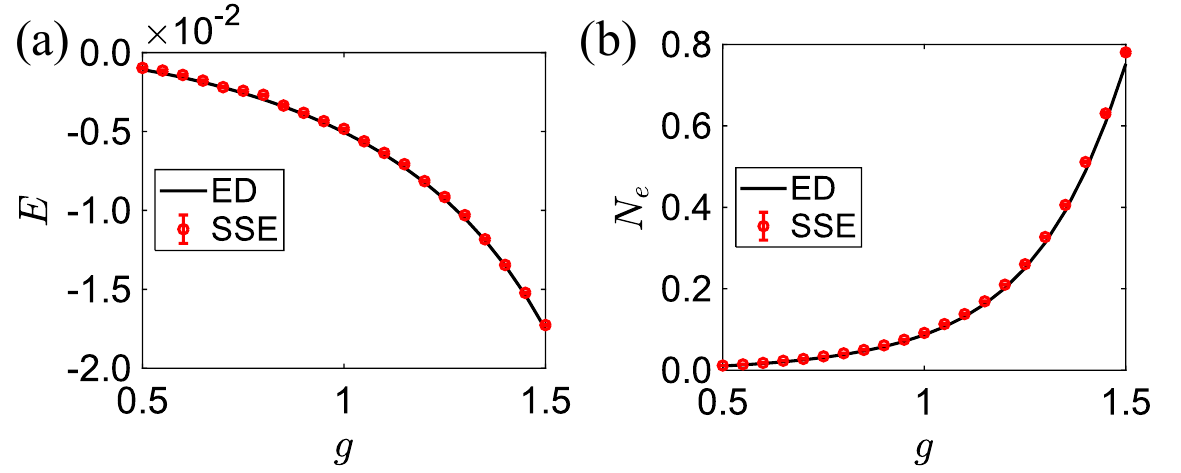}
    \caption{Comparison of (a) energy per site and (b) number of excited particles with SSE and ED methods. The parameters of the system are chosen as $\mu = -3.4$, $\Delta = 3$, $\alpha = 0.5$, and $V=1$. Meanwhile, the photon number is truncated to $n_{\mathrm{p}} = 8$ in the ED. }
    \label{CED}
\end{figure}

From the above discussion, and including the case when the vertex remains unchanged (a special type of bounce), we see that there are up to six different types of updates for some given vertex. 
To satisfy detailed balance, the acceptance probabilities for these operator-update schemes must be determined.
Therefore, we extend the optimized transfer matrix introduced by Ref.~\cite{Sylju_sen_2002} to arbitrarily large dimensions; see Appendix \ref{Loop} for details. Meanwhile, similar to our previous algorithm \cite{zhang01}, there is no truncation error related to the photon field in our SSE algorithm.

We have benchmarked this SSE algorithm against zero-temperature exact diagonalization (ED). The inverse temperature is set to $\beta=200$, which is much smaller than the other energy scales. The energy per site and the number of excited particles, given by $N_{e} = \langle a^{\dagger} a + \sum_{j = 1}^{N} n_{j} \rangle$, are calculated via SSE and compared with the ED results on a square lattice with $N = 16$ sites. As shown in Fig.~\ref{CED}, the two sets of results are in excellent agreement, thereby validating the SSE method.
We expect that the development of robust, large-scale SSE simulations will allow us to benchmark future cavity QED experiments based on the Rydberg array platform. Furthermore, this algorithm is also suitable for the circuit-QED system. In the rest of this article, we will focus on the SRPT (associated with the spontaneous breaking of the $Z_{2}$ symmetry) and other interesting properties of the model in the square lattice.

\section{Numerical Results without Rydberg Interaction}
\label{numerical1}

The Ising interaction between Rydberg atoms is of the van der Waals type, so it decays as the sixth power of distance and can be ignored when the distance between atoms is sufficiently large. In this limit, Eq.~\eqref{Ohamil} reduces to the ADM. The number of photons $n_{\rm ph} = \langle a^{\dagger}a \rangle$ can be chosen as an order parameter for the quantum phase transition of the ADM: In the SR phase, the photons and atomic excitations form polaritons, leading to a finite occupation $n_{\rm ph}$ of the cavity mode.              
From the point of view of symmetry, we note that ADM remains invariant under the transformation $(a, a^{\dagger}) \to (-a, -a^{\dagger})$, $(\sigma_{j}^{-}, \sigma_{j}^{+}) \to (-\sigma_{j}^{-}, -\sigma_{j}^{+})$, and $\sigma_{j}^{z} \to \sigma_{j}^{z}$, which is commonly referred to as $Z_{2}$ symmetry.
The corresponding conserved quantity, known as parity, is defined as $P = (-1)^{a^{\dagger} a}\prod_{i=1}^{N} \sigma_{i}^{z}$. 
Owing to the commutation relation $[P, H] = 0$, the $P$ operator and the system Hamiltonian possess a complete set of common eigenstates.
Each eigenstate can be labeled by its even or odd parity, corresponding to the eigenvalues $+1$ or $-1$ of $P$, respectively.
While the ground state of NP possesses an even parity $P = +1$, the ground state of the SR phase becomes degenerate in the thermodynamic limit, and we can construct a superposition mixing different parities, giving $\langle P \rangle = 0$.
Parity, therefore, provides a clear criterion for distinguishing the normal and superradiant phases \cite{PhysRevE.67.066203}.
Furthermore, the $Z_{2}$ invariance is promoted to a $U(1)$ symmetry when $\alpha = 0$.
More specifically, the system becomes invariant under the transformation $(a, a^{\dagger}) \to (ae^{-i\theta}, a^{\dagger}e^{i\theta})$, $(\sigma_{j}^{-}, \sigma_{j}^{+}) \to (\sigma_{j}^{-} e^{-i\theta}, \sigma_{j}^{+} e^{i\theta})$ and $\sigma_{j}^{z} \to \sigma_{j}^{z}$.
Therefore, the total number of excitations $N_{e}$ is the conserved quantity associated with the $U(1)$ symmetry.

\begin{figure}
    \centering
    \includegraphics[width=1.0\linewidth]{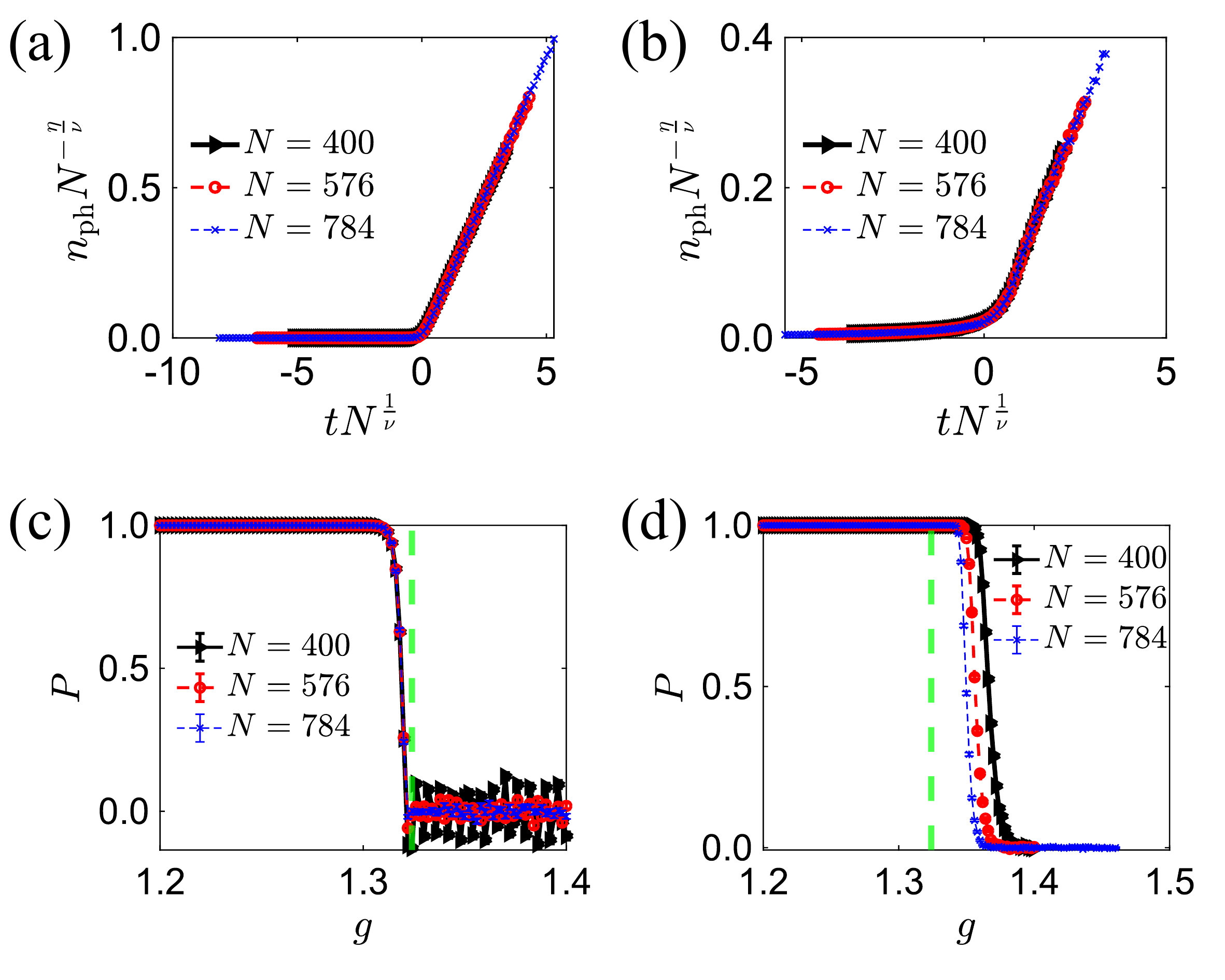}
    \caption{Dependence of various physical quantities on the coupling strength $g$ around the phase transition point. (a,b) are FSS analysis of $n_{\rm ph}$, and (c, d) present the parity as a function of $g$. Here, the green vertical dashed lines indicate the critical point obtained through FSS, see Table~\ref{ADMcritE}. The parameters are $\mu = -3.5$ and $\Delta = 3$, with the system sizes $N = 400$ (black triangles), $576$ (red circles), and $784$ (blue crosses), for (a,c) $\alpha=0$ and (b,d) $\alpha=0.5$.}
    \label{Scaling}
\end{figure}

Due to the exponential growth of the Hilbert space with system size, ED becomes computationally prohibitive for systems as small as $N \approx 30$. Therefore, to extract the critical behavior of the SR phase transition, we rely on SSE simulations. Near the quantum critical point, the correlation length diverges as $\xi \sim |t|^{-\nu}$, where the reduced coupling strength is defined as $t=\frac{g-g_{c}}{g_{c}}$, and the photon number is a singular but non-divergent order parameter, with the dependence $n_{\mathrm{ph}} \sim |t|^{-\eta}$. After obtaining the desired physical quantities at various system sizes, we perform a finite-size scaling (FSS) analysis of the numerical data by considering the dependence of $y_{N}=n_{\rm ph} N^{-\frac{\eta}{\nu}}$ on $x_{N} = t N^{\frac{1}{\nu}}$, which follows a well-defined scaling law. Since the critical exponent of the correlation length is universal, as confirmed by other analytical and numerical results~\cite{LiuTao, Lambert073602, JVidal_2006}, we impose $\nu = 3/2$ and determine $g_c$ and $\eta$ through the method proposed by Houdayer and Hartmann, i.e., by minimizing the loss function $S$ defined by Eq.~(A1) of Ref.~\cite{Houdayer014418}. We generally obtain an excellent data collapse, as shown in Fig.~\ref{Scaling}(a) and (b) for  $\alpha=0$ and~$0.5$, respectively.

\begin{table}
    \centering
    \caption{SRPT critical points $g_{c}$ and photon critical exponents $\eta$ at different values of $\alpha$, for $\mu =-3.5$ and $\Delta = 3$.}
    \begin{tabular}{cccccc}
        \hline \hline
        $\alpha$ & $0$ & $0.1$ & $0.5$ & $0.9$ & $1.0$   \\
        \hline
        $\eta$ & $0.507$ & $0.521$ & $0.522$ & $0.494$ & $0.464$  \\
        Error ($\eta$) & $\pm 0.091$ & $\pm 0.061$ & $\pm 0.051$ & $\pm 0.044$ & $\pm 0.088$ \\
        $g_{c}$ & $1.322$ & $1.323$ & $1.323$ & $1.323$ & $1.322$ \\
        Error ($g_{c}$) & $\pm 0.001$ & $\pm 0.003$ & $\pm 0.003$ & $\pm 0.002$ & $\pm 0.003$ \\
        \hline \hline
    \end{tabular}
    \label{ADMcritE}
\end{table}

Estimates of the critical coupling $g_c$ and critical exponent $\eta$ at different values of $\alpha$ are reported in Table~\ref{ADMcritE}, with the corresponding errors.
We find that the critical exponents converge to nearly identical values in the thermodynamic limit, compatible with $\eta=1/2$, despite obvious differences in the finite-size photon behavior around the critical point.
Therefore, we conclude that the photon number follows a common scaling law in the critical regime.
Table~\ref{ADMcritE} also shows that the critical coupling $g_c$ is independent of $\alpha$ within error bars and equals the analytic solution $g_c=\sqrt{\mu(\mu+\Delta)}\approx1.323$ \cite{PhysRevE.67.066203}. These results strongly support the analytic conclusion that the critical point and scaling behavior of the SR phase transition are independent of the NAP \cite{PhysRevA.8.1440}.

On the other hand, the interplay between the RW and CRW terms leads to finite-size effects that depend strongly on $\alpha$. This is most evident in the behavior of the ground-state parity, shown in (c) and (d) of Fig.~\ref{Scaling} for various representative cases. In Fig.~\ref{Scaling}(c), where $\alpha=0$, the parity vanishes almost simultaneously at the critical point, for all system sizes. A similar behavior is found at $\alpha=1$. In these two cases, the quantum phase transition breaks the $U(1)$ symmetry and $N_{e}$ acquires a finite value in the SR phase. As also confirmed by the ED results presented in Appendix~\ref{ED}, see in particular Fig.~\ref{EDparity}(a), increasing $g$ causes the ground-state value of $N_{e}$ to grow in steps, with each integer corresponding to a specific parity. This behavior is still captured by our SSE simulations, which are performed at low but finite temperatures. At small values of $N$, the energy gap between the ground and first excited state is sufficiently large, making the contribution of the finite-size ground state dominant. As a consequence, the numerical data of Fig.~\ref{Scaling}(c) show strong oscillations in the SR phase when $N=400$. At larger system sizes, the low-energy states become approximately degenerate, and these oscillations eventually disappear. However, it is expected to observe such oscillations in the real experiment due to the limitation of space in the cavity.

Interesting, a different behavior of the parity is found at intermediate values of the NAP. We consider in Fig.~\ref{Scaling}(d) the case $\alpha=0.5$, where no oscillations are seen, and the coupling strength required to make the ground-state parity vanish is significantly larger than the critical value $g_{c}$. However, we can clearly find the finite size transition point of the parities approaching $g_{c}$ while increasing the system size. In the analytic treatment \cite{PhysRevE.67.066203}, the ensemble of two-level atoms can be taken as a pseudospin of length $J=N/2$. Then, after the Holstein-Primakoff transformation, the Hamiltonian can be transformed to an effective two-mode bosonic Hamiltonian, which can be exactly solved in the thermodynamic limit. However, in the finite system, the non-linear terms $\sim O(1/N)$ due to the Holstein-Primakoff transformation can not be omitted. Considering the parity is very sensitive to the change of $N_e$, reflected in Fig.~\ref{Scaling}, it suffers from a more serious finite-size effect than the photon density. Therefore, the different behaviors of the parities at different NAPs exhibit distinctive features beyond the mean-field level which will be strongly enhanced when including the Ising interactions.

\section{Numerical Results With Rydberg interaction}
\label{numerical2}

The experimental platform based on Rydberg atoms is dominated by the strong and controllable interaction induced by the large dipole moments of the highly-excited Rydberg states. By expressing it through the occupation numbers of the Rydberg states, it is described as an effective Ising interaction. Because the corresponding repulsive interaction is of van der Waals type, we consider only nearest-neighbor interactions here. Different from the ADM, as a strongly correlated system, the Hamiltonian Eq.\ref{Ohamil} can not be exactly solved, so the numerical simulation becomes necessary. To detect the spontaneous translational symmetry breaking on the square lattice, we can utilize the static structure factor $S({\bf q})/N =\langle |\sum_{j=1}^{N} n_{j} e^{i{\bf q} \cdot {\bf r}_{j}}|^{2} \rangle /N^{2} $ as an order parameter. The wavevector ${\bf q}$ is set to $(\pi,\pi)$, as appropriate for an antiferromagnetic interaction $V > 0$. In fact, $S(\pi, \pi)/N$ is equivalent to the staggered magnetization and tends to a finite value both in the Solid-1/2 and SRS phases. The corresponding Binder cumulant is defined as $S_{b} = \frac{3}{2}(1 - \frac{1}{3} \frac{\langle S^{2}({\bf q}) \rangle}{\langle S({\bf q}) \rangle ^{2}})$. According to the mean-field phase diagram in Fig.\ref{ADMdata}, we analyze the effect of NAP $\alpha$ on the quantum phase transition among NP, SR, SRS, and solid-1/2 phases in the following.

\begin{figure}
    \centering
    \includegraphics[width=1.0\linewidth]{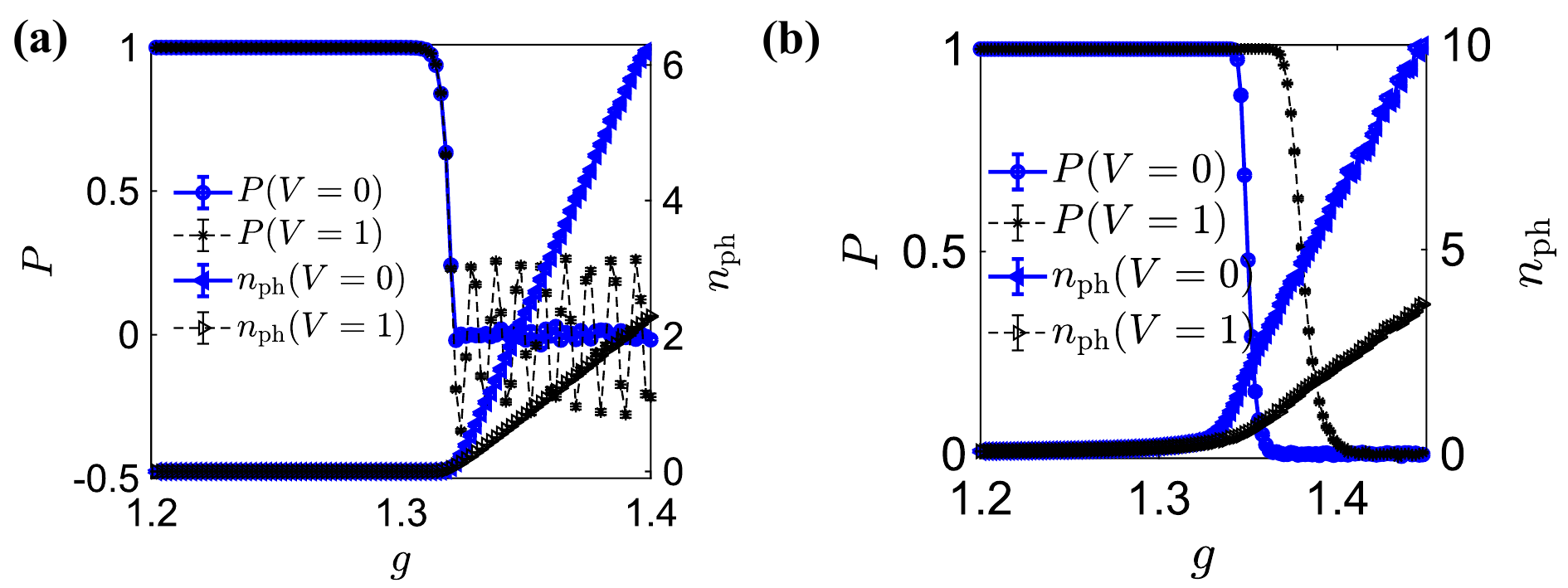}
    \caption{Effects of the Rydberg-Rydberg interaction on the parities and the number of photons during the SR phase transition for (a) $\alpha=0$ and (b) $\alpha=0.5$, respectively. Other parameters are $\mu=-3.3$, $\Delta=3$, and $N = 784$.
    }
    \label{phsupress}
\end{figure}
 

First, we investigate the transition from the NP to the SR phase in the presence of the Ising term on a square lattice. The large-scale SSE results of Fig.~\ref{phsupress} demonstrate that the Ising interaction does not modify the boundary of this phase transition in the thermodynamic limit. However, the presence of the Rydberg interaction can strongly change the behavior of the observables. In the SR phase, a significant reduction of $n_{\rm ph}$ at different NAP can be clearly observed, which indicates the excitation of the polariton becomes harder due to the strong Rydberg repulsive interactions. On the other hand, the parity appears more sensitive. At the $U(1)$ symmetry point $\alpha=0$ in Fig.\ref{phsupress}(a), the Rydberg interaction strongly boosts the oscillation, which hints that the corresponding gap energies between the even and odd parity sectors are enlarged. When turning on the CRW term at $\alpha=0.5$, the Fig.\ref{phsupress}(b) shows that the vanishing of $P$ is pushed to larger $g$ by the Rydberg interaction, which indicates the prefactor of the finite size correction is enhanced while increasing $V$.

\begin{figure}
    \centering
    \includegraphics[width=1.0\linewidth]{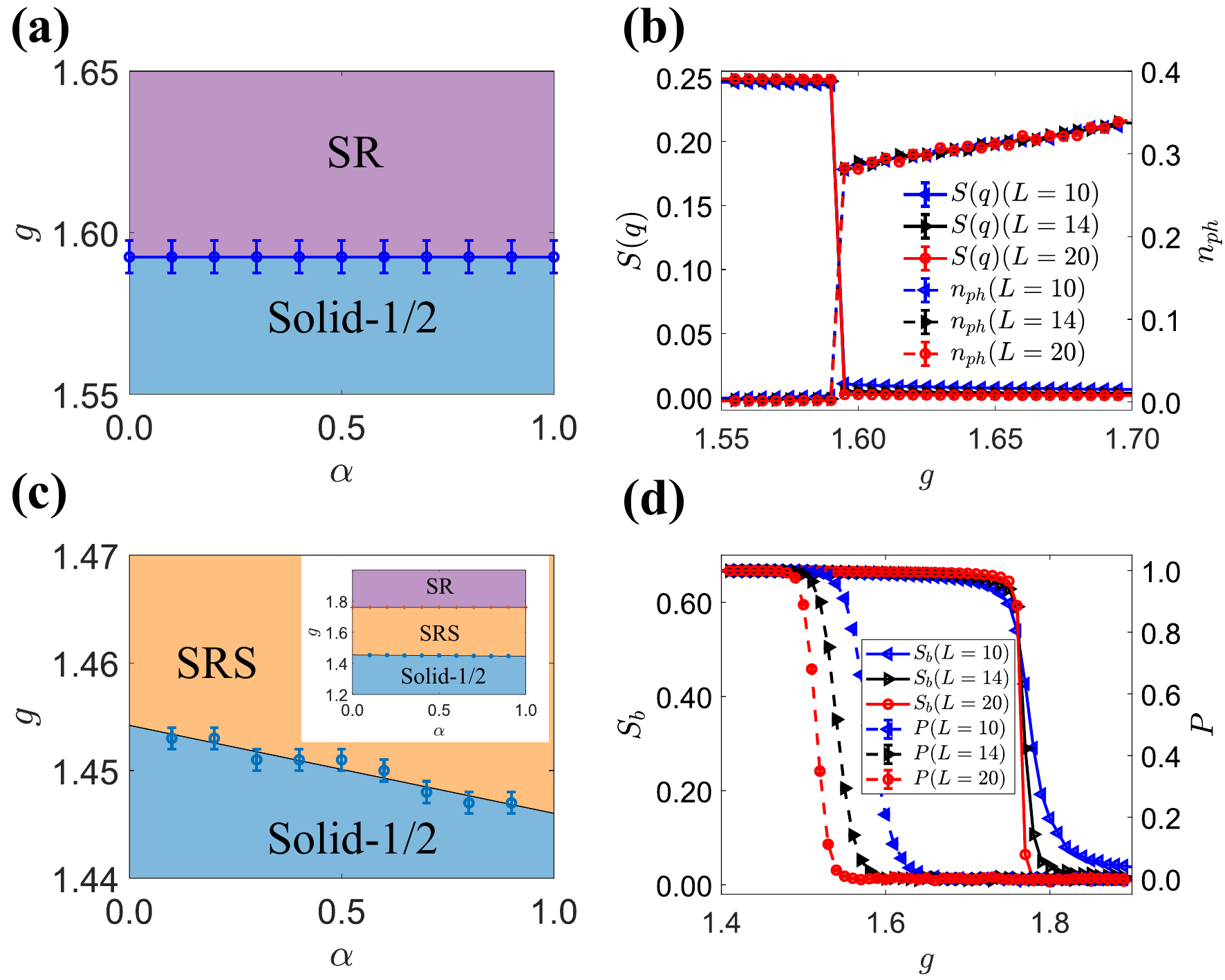}
    \caption{(a) The first-order phase transition line between the Solid-1/2 and SR phases at $\mu = -1.4$. 
    (b) $S(q)$ and $n_{\mathrm{ph}}$ as functions of $g$ at $\alpha = 0.5$ with different system sizes. 
    (c) The second-order phase transition line between the Solid-1/2 and SRS phases at $\mu = -2.5$, and critical lines related to the SRS phase are shown in the inset. (d) Binder cumulant $S_{b}$ and parity $P$, plotted as functions of $g$  for various system sizes and $\alpha = 0.5$. All data were obtained with $\Delta = 3$ and $V = 1$.
    }
    \label{solid}
\end{figure}

The interplay of RW, CRW, and Ising terms yields a rich phase diagram that includes solid phases.  We have explored them through SSE simulations in Fig.~\ref{solid}, finding that the phase transition from the Solid-1/2 to the SR phase, which takes place at small values of $\mu$, exhibits remarkable robustness. Figure~\ref{solid}(a) demonstrates that the critical point (blue solid dots) is also independent of $\alpha \in [0,1]$. The transition from the Solid-1/2 to the SR phase, which
restores translational symmetry while simultaneously breaking the $Z_{2}$ symmetry, remains first-order at arbitrary values of $\alpha$.
This is confirmed in Fig.~\ref{solid}(b), where $\alpha=0.5$. At the critical point, the discontinuous vanishing of the order parameter $S(q)$ is accompanied by a jump of $n_{\mathrm{ph}}$, which acquires a finite value in the SR phase.

In our previous work \cite{zhang01,AnGaoqi,liang2025}, the most fascinating phenomena would be the SRS phase as an intermediate phase between SR and solid-1/2 phases. Differently, when including the CRW interaction, the SRS phase spontaneously breaks both translational and $Z_{2}$ symmetries. The quantum phase diagram from Solid-1/2 to SRS, and finally to the SR phase, is shown as a function of $\alpha$ in the inset of Fig.~\ref{solid}(c). For all values of $\alpha$, the former transition is second-order, while the latter is first-order.
The exact phase transition points are determined by collapsing the parity $P$ and the Binder cumulant $S_{b}$, obtained at different system sizes. For example, Fig.~\ref{solid}(d) displays these quantities at $\alpha = 0.5$.
Same as the phase transition between solid-1/2 and SR phase with the disappearance of solid order, the SRS and SR phases also undergo the first-order transition, and meanwhile, the phase boundary between the SRS and SR phases is also unaffected by $\alpha$. However, the critical coupling between Solid-1/2 and SRS phases shows a small linear shift with $\alpha$ which can be seen in Fig.~\ref{solid}(c).  The corresponding melting of the Solid-1/2 phase is expected to result from the energy gap closing of inserting an additional quasi-particle. However, the effect of the RW and CRW interaction is to couple a photon with a hole $|\downarrow\rangle$ or particle $|\uparrow\rangle$, respectively. Considering the particle-hole or spin up-down symmetry is artificially broken, the interplay between RW and CRW terms can strongly change the critical line between SRS and Solid-1/2 phases.


\section{Conclusion and Outlook}
\label{outlook}
In conclusion, we have numerically investigated the ADIM with a cluster SSE algorithm specifically designed for this model. 
Our key findings are as follows: 
First, we have benchmarked the scaling law of the photon number at the SR critical point of the ADM and observed the vanishing of parity in finite-size systems when entering the SR phase. 
Second, for the full ADIM, we have demonstrated that the Rydberg-Rydberg interactions suppress the order parameter of the SR phase and that quantum fluctuations from the CRW terms shift the second-order phase boundary between the SRS and Solid-1/2 phases. 
Finally, we have confirmed that both transitions from the Solid-1/2 or SRS to the SR phase are first-order, for any value of the NAP in the square lattice.

Our work paves the way for several future research directions. 
The interplay of anisotropic coupling with geometrical frustration in different lattice geometries is a promising avenue for discovering new phases of quantum matter. 
Furthermore, our improved SSE algorithm is readily applicable to a broader class of light-matter systems where CRW terms play an essential role. Following a recent proposal for realizing the ADIM with Rydberg atom arrays~\cite{dongade388}, we anticipate direct tests of our predictions based on the cavity-Rydberg platforms. Meanwhile, the corresponding phenomena may also be checked in the circurt-QED platform.


\acknowledgements
X.-F. Z. acknowledges funding from the National Science Foundation of China under Grants  No. 12274046, No. 11874094, No.12147102, and No.12347101, Chongqing Natural Science Foundation under Grants No. CSTB2022NSCQ-JQX0018, Fundamental Research Funds for the Central Universities Grant No. 2021CDJZYJH-003, and Xiaomi Foundation / Xiaomi Young Talents Program.
S.C. acknowledges support from the Innovation Program for Quantum Science and Technology (Grant No. 2021ZD0301602) and the National Science Association Funds (Grant No. U2230402).

\appendix
\section{Redesigned Loop Update}
\label{Loop}

To improve the efficiency of the algorithm, the acceptance probability must be minimized, where the vertex is not updated (here, the bounced-back process can also update the vertex). 
Therefore, we discuss the general case of the number of possible transferred vertices $m > 4$,  and define $P(s \to s') = a_{s, s'}/W_{s}$, where $P$ is the probability that the vertex $s$ transfers to the vertex $s'$, $W_{s}$ represents the weight of the vertex $s$, and $W_{1} \ge W_{2} \ge ... \ge W_{m}$, $\delta = -W_{1} + W_{2} + W_{3} + W_{4}$.
Then, the detailed balance condition re-expresses as $a_{s, s'} = a_{s', s}$ and $\sum_{s'} a_{s, s'} = W_{s}$.
Following the Ref.\cite{Sylju_sen_2002}, for the condition $W_{1} \ge W_{2} + W_{3} + ... + W_{m}$, one optimal solution is very simple as
\begin{align}
    &a_{1, 1} = W_{1} - \sum_{k=2}^{m}W_{k} \nonumber\\
    &a_{1, 2} = W_{2}\nonumber\\
    &... \nonumber\\
    &a_{1, m} = W_{m}
\end{align}
Then if $W_1<\sum_{i>1}^mW_i$ but $\delta > 0$, one possible optimal solution with diagonal terms $a_{i,i}$ vanished is
\begin{align}
    &a_{1,2} = \left( W_{1} + W_{2} - W_{3} - W_{4}\right) / 2 \nonumber\\
    &a_{1,3} = \left( W_{1} - W_{2} + W_{3} - W_{4}\right) / 2 \nonumber\\
    &a_{2,3} = \left( - W_{1} + W_{2} + W_{3} + W_{4}\right) / 2 \nonumber\\
    &a_{1,4} = W_{4} - W_{5} / 2 \nonumber\\
    &a_{1,5} = W_{5} - W_{6} / 2 \nonumber\\
    &... \nonumber\\
    &a_{1,m-1} = \left( W_{m-1} - W_{m} \right) / 2 \nonumber\\
    &a_{1,m} = W_{m} / 2 \nonumber\\
    &a_{4,5} = W_{5} / 2 \nonumber\\
    &a_{5,6} = W_{6} / 2 \nonumber\\
    &a_{m-1, m} = W_{m} / 2
    \label{sedbq}
\end{align}
Beyond Ref.\cite{Sylju_sen_2002}, we demonstrate that, for $\delta < 0$, one can also get a similar optimal solution with zero diagonal terms.
Because of $W_{1} < W_{2} + W_{3} + ... + W_{m}$, there always exists a number $k$ satisfying $\sum_{l=k+1}^{m} W_{l} < - \delta < \sum_{l=k}^{m}W_{l}$.
Then we choose $a_{1, l} = a_{l, 1} = W_{l}$ and other elements equal to zeros for $l > k$.
After introducing $\delta' = -\delta - \sum_{l = k + 1}^{m} W_{l}$ and  $a'_{1, k} = a_{1, k} - \delta'$, we obtain new detailed balance equations with fewer conditions $k\le m$
\begin{align}
    &\sum_{l=1}^{k-1} a_{1, l} + a'_{1, k} = W_{1} + \delta = W'_{1} \nonumber\\
    &\sum_{l=1}^{k} a_{2, l} = W_{2} \nonumber\\
    &\sum_{l=1}^{k} a_{3, i} = W_{3} \nonumber\\
    & ... \nonumber\\
    & \sum_{l=1}^{k} a_{k-1, l} = W_{k-1} \nonumber\\
    &a'_{1, k} + \sum_{l=2}^{k} a_{k, i} = W_{k} - \delta' = W'_{k} > 0
    \label{thrid}
\end{align}
Due to $W'_{1} - W_{2} - W_{3} - W_{4} = 0$, the general solution of the second part Eq.\eqref{sedbq} is also suitable for solving Eq.\eqref{thrid}.
We stress the final transfer probability $P(s \to s') = a_{s, s'} / W_{s}$, and $a_{1, k} = a'_{1, k} + \delta$.
In addition to the mentioned $a_{s, s'}$ and its conjugate, all other terms are zero.
Thus, a complete directed loop involves: randomly selecting a leg of a vertex, recording the operation (add or minus one particle for bosons, flip up or down for atoms) on this leg, and then updating according to the solutions above. 
The process continues until the worm head returns to the initial leg with an operation opposite to the start.

\section{Exact Diagonalization Results}
\label{ED}

\begin{figure}
	\centering
	\includegraphics[width=1.0\linewidth]{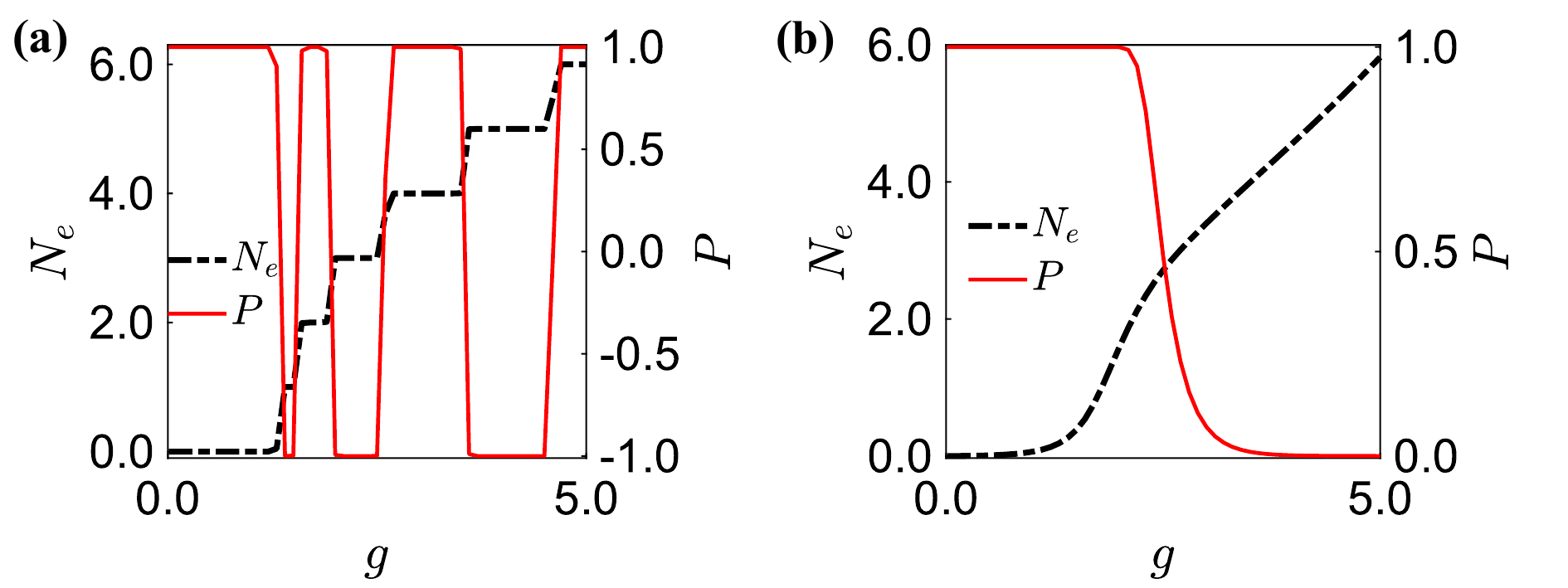}
	\caption{ADM results are obtained by ED: the number of excited particles (black dotted line) and the parity (red solid line) of ground state with (a) $\alpha = 0$ and (b) $\alpha = 0.5$ from NP to SRP, which is considered in 1D chain $N = L = 6$, $\Delta = 3$, $\mu = -3.5$, and the number of photon truncation is $n_{\mathrm{p}} = 10$. }
	\label{EDparity}
\end{figure}

In the QRM or few-body ADM, the interplay between the CRW and RW gives rise to a host of intriguing phenomena, which become particularly evident when comparing the cases before and after taking the RWA.
We also surveyed the ground state properties via the zero-temperature ED for a small system size.
In Fig. \ref{EDparity}, we compare $P$ and $N_{e}$ for $\alpha = 0, 0.5$, with all other parameters held fixed.
For $\alpha = 0$, each $N_{e}$ plateau corresponds to a specific parity, resulting in parity oscillations in the ground state, see Fig. \ref{EDparity}(a).
However, for $\alpha = 0.5$, $N_{e}$ increases smoothly with $g$.
In the SR region, parity breaking occurs due to the ground-state degeneracy between even and odd parity, as shown in Fig. \ref{EDparity}(b).
Finally, we note that as the system size increases, the atom-field coupling strength is enhanced by a factor of $\sqrt{N}$, causing the behavior in Fig. \ref{EDparity}(a) to approach that shown in Fig. \ref{EDparity}(b).

\bibliographystyle{apsrev4-1}
\bibliography{referen}

\end{document}